\documentclass[letterpaper,twocolumn,10pt]{article}
\usepackage{usenix,epsfig,endnotes}

\usepackage{times}
\usepackage{listings}
\usepackage{textcomp}
\usepackage[T1]{fontenc}
\PassOptionsToPackage{hyphens}{url}\usepackage[colorlinks=true,urlcolor=blue,citecolor=blue,linkcolor=blue,plainpages=false]{hyperref}
\usepackage[normalem, normalbf]{ulem}
\usepackage{flushend}
\usepackage{xcolor}
\usepackage{multirow}
\usepackage{booktabs}
\usepackage{graphicx}
\usepackage{enumitem}
\usepackage{changes}
\usepackage{amsmath}
\usepackage[TABBOTCAP]{subfigure} \subfigcapmargin = 0.5cm
\usepackage{xspace}

\newcommand{\hiddenoutline}[1]{}

\newcommand{\name}[0]{Nimbus\xspace}

\begin{document}

%don't want date printed
\date{}

%make title bold and 14 pt font (Latex default is non-bold, 16 pt)
% \title{\Large \bf Execution Templates: Flexible Scheduling for High Task Throughput Jobs}
\title{\Large \bf Execution Templates: Caching Control Plane Decisions for
Strong Scaling of Data Analytics}

%for single author (just remove % characters)
\author{
{\rm Omid Mashayekhi}\\
Stanford University
\and
{\rm Hang Qu}\\
Stanford University
\and
{\rm Chinmayee Shah}\\
Stanford University
\and
{\rm Philip Levis}\\
Stanford University
% copy the following lines to add more authors
% \and
% {\rm Name}\\
%Name Institution
} % end author

\maketitle

% Use the following at camera-ready time to suppress page numbers.
% Comment it out when you first submit the paper for review.
\thispagestyle{empty}

\subsection*{Abstract}
Control planes of cloud frameworks trade off between scheduling granularity
and performance. Centralized systems schedule at task granularity, but
only schedule a few thousand tasks per second. Distributed systems
schedule hundreds of thousands of tasks per second but changing the
schedule is costly.

We present {\it execution templates}, a control plane abstraction that can
schedule hundreds of thousands of tasks per second while supporting
fine-grained, per-task scheduling decisions. Execution templates leverage a
program's repetitive control flow to cache blocks of frequently-executed tasks.
Executing a task in a template requires sending a single message. Large-scale
scheduling changes install new templates, while small changes apply edits to
existing templates.

Evaluations of execution templates in \name, a data analytics framework, find
that they provide the fine-grained scheduling flexibility of centralized
control planes while matching the strong scaling of distributed ones.
Execution templates support complex, real-world applications, such as a fluid
simulation with a triply nested loop and data dependent branches.

\section{Introduction}

As data analytics have transitioned from file I/O~\cite{hadoop,mapreduce} to
in-memory processing~\cite{graphlab,naiad,spark}, systems have focused on
optimizing the CPU performance~\cite{ousterhout15}.  Spark 2.0, for example,
reports 10x speedups over prior versions with new code generation
layers~\cite{databricks-spark2}.  Introducing data-parallel optimizations such
as vectorization, branch flattening, and prediction can in some cases be faster
than hand-written C~\cite{nvl-mit, nvl-platformlab}. GPU-based
computations~\cite{torch,tensorflow} improve performance further.

\begin{figure}
\centering
\includegraphics[width=3.0in]{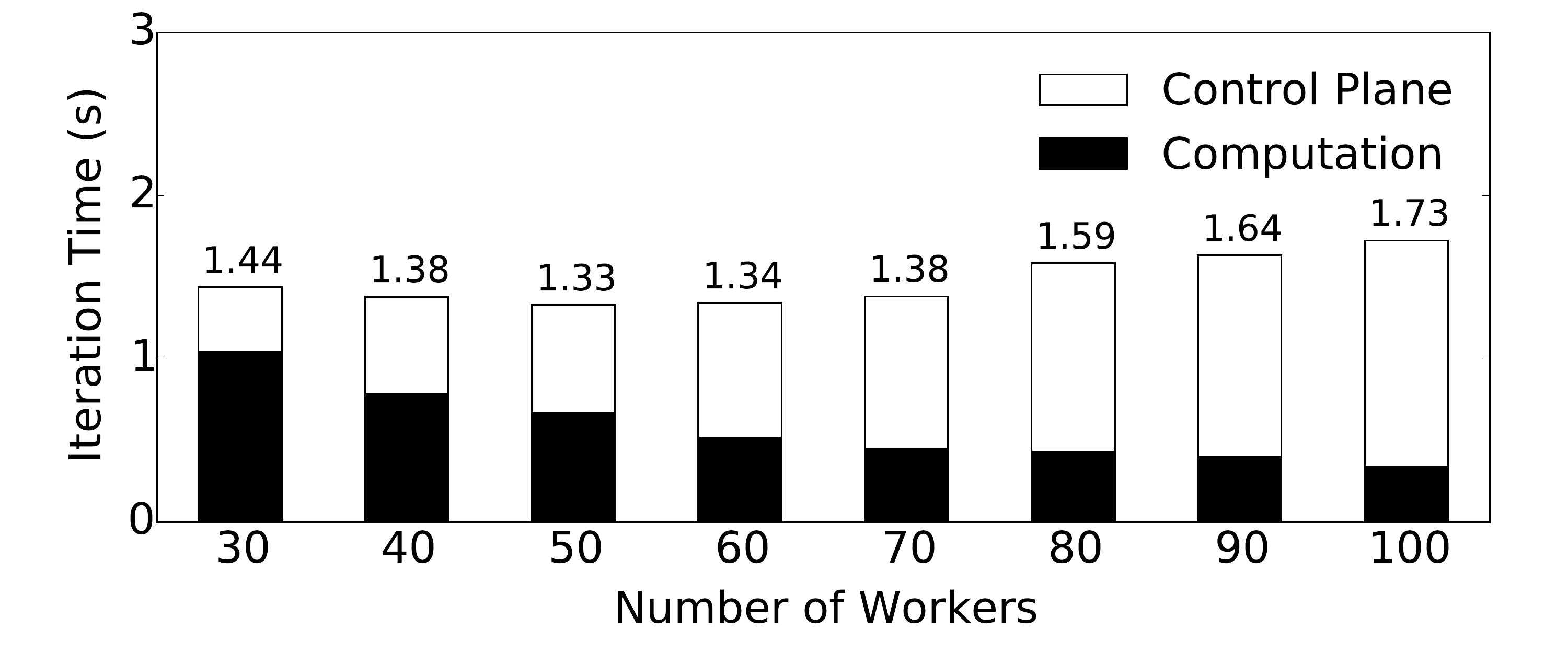}
\caption{The control plane is a bottleneck in modern analytics
  workloads. Increasingly parallelizing logistic regression on 100GB
  of data with Spark 2.0's MLlib reduces computation time (black bars)
  but control overhead outstrip these gains, {\it increasing}
  completion time.}
\label{fig:spark-mllib-strong}
\end{figure}

Speedup in computations, however, demands a higher task throughput from the
control plane. This creates new tension between task throughput requirements at
scale on one hand, and dynamic, fine-grained scheduling decisions on the other.
Available systems cannot fulfill both requirements simultaneously. Today,
frameworks adopt one of two design points to schedule their computations across
workers. One is a centralized controller model, and the other is a distributed
data flow model.

In the first model, systems such as Spark~\cite{spark} use
a centralized control plane, with a single node that dispatches small
computations to worker nodes. Centralization allows a framework to quickly
reschedule, respond to faults, and mitigate stragglers {\it reactively}, but as
CPU performance improves the control plane becomes a bottleneck.
Figure~\ref{fig:spark-mllib-strong} shows the performance of Spark 2.0's MLlib
logistic regression running on 30--100 workers.  While computation time
decreases with more workers, these improvements do not reduce overall
completion time. Spark spends more time in the control plane, spawning and
scheduling computations.
While there is a huge body of work for scheduling {\it multiple} jobs within a
cluster~\cite{apollo,hawk,tarcil,mesos,mercury,sparrow,yarn}, these approaches
do not help when a {\it single} job has a higher task throughput than what the
control plane can handle, as in Figure~\ref{fig:spark-mllib-strong}.

The second model, used by systems such as Naiad~\cite{naiad} and
TensorFlow~\cite{tensorflow}, is to use a fully distributed control plane. When
a job starts, these systems install data flow graphs on each node, which then
independently execute and exchange data. By distributing the control plane and
turning it into data flow, these frameworks achieve strong scalability at hundreds of thousands
of tasks per second.  However, data flow graphs describe a static schedule.
Even small changes, such as migrating a task between two nodes, requires
stopping the job, recompiling the flow graph and reinstalling it on every node.
As a result, in practice, these systems mitigate stragglers only {\it
proactively} by launching backup workers, which requires
extra resource allocation even for non-straggling tasks~\cite{tensorflow}.

This paper presents a new point in the design space, an abstraction called
{\it execution templates}. Execution templates schedule at the same per-task
granularity as centralized schedulers.  They do so while imposing the same
minimal control overhead as distributed execution plans.  Execution templates
leverage the fact that long-running jobs (e.g. machine learning, graph
processing) are repetitive, running the same computation many
times~\cite{ernest}.
% Machine learning algorithms, for example, typically
% iterate until the estimation error drops below a threshold.

Logically, a framework using execution templates centrally schedules at task
granularity. As it generates and schedules tasks, however, the system caches
its decisions and state in templates. The next time the job reaches the same
part of its program, the system executes from the templates rather than resend
all of the tasks. Depending on how much system state has changed since the
template was installed, a controller can immediately {\it instantiate} the
template (i.e.  execute without modification), {\it edit} the template by
changing some of its tasks, or {\it install} a new version of template.
Templates are not bound to a static control flow and support data-dependent
branches; controllers {\it patch} system state dynamically at runtime if
needed.  We call this abstraction a template because it caches some information
(e.g., dependencies) but instantiation requires parameters (e.g., task IDs).

Using execution templates, a centralized controller can generate and
schedule hundreds of thousands of low-latency tasks per second. 
We have implemented execution templates in \name, an analytics
framework designed to support high performance computations.
This paper makes five contributions:

\begin{enumerate}[leftmargin=*] % [nolistsep, leftmargin=*]
  \setlength{\itemsep}{.5pt}

\item Execution templates, a control plane abstraction that schedules
  high task throughput jobs at task granularity (Section~\ref{sec:templates}).

\item A definition of the requirements execution templates place on a
  control plane and the design of \name, a framework that meets these
  requirements (Section~\ref{sec:design}).

\item Details on how execution templates are implemented in \name, 
  including
  program analyses to generate and install efficient templates,
  validation and patching templates to meet their preconditions, and
  dynamic edits for in-place template changes
  (Section~\ref{sec:implementation}).

\item An evaluation of execution templates on analytics benchmarks,
  comparing them with Spark's fine-grained scheduler and Naiad's
  high-throughput data flow graphs (Section~\ref{sec:evaluation}).

\item An evaluation of \name running a PhysBAM~\cite{physbam} particle-levelset
  water simulation~\cite{particle-levelset} with tasks as short as 100$\mu$s. 
  (Section~\ref{sec:evaluation}).\footnote{PhysBAM
  is an open-source simulation package that has received two Academy Awards 
  and has been used in over 20 feature films.}

\end{enumerate}

This paper does not examine the question of scheduling policy, e.g.,
how to best place tasks on nodes, whether by min-cost flow
computations~\cite{firmament,quincy}, packing~\cite{tetris,graphene},
or other algorithms~\cite{apollo,mesos,mercury,omega}
(Section~\ref{sec:related}). Instead, it looks at the {\it mechanism}:
how can a control plane support high throughput, fine-grained
decisions?  Section~\ref{sec:discuss} discusses how execution
templates can be integrated into existing systems and concludes.

\section{Execution Templates}

\label{sec:templates}

\begin{figure}
    \centering
    \includegraphics[width=3.0in]{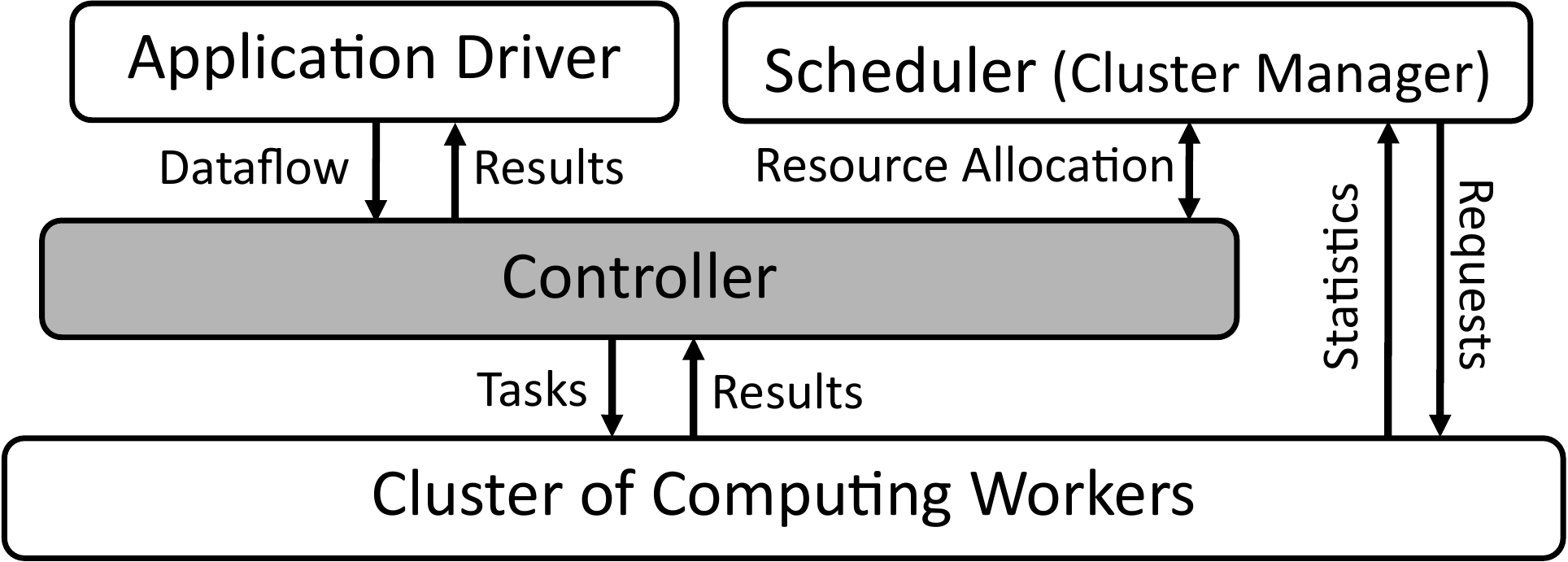}
    \caption{Generalized architecture of a cloud computing system: a
      {\it driver} program specifies the application logic to a {\it
        controller}, which can either directly assign tasks to workers
      or request resources from a {\it cluster manager}. Execution
      templates operate within a controller.}
    \label{fig:architecture}
\end{figure}

\begin{figure*}
\centering
\setlength\tabcolsep{-3pt}
\begin{tabular}{cc}
\subfigure[Driver program pseudocode.]
{
\includegraphics[width=3.0in]{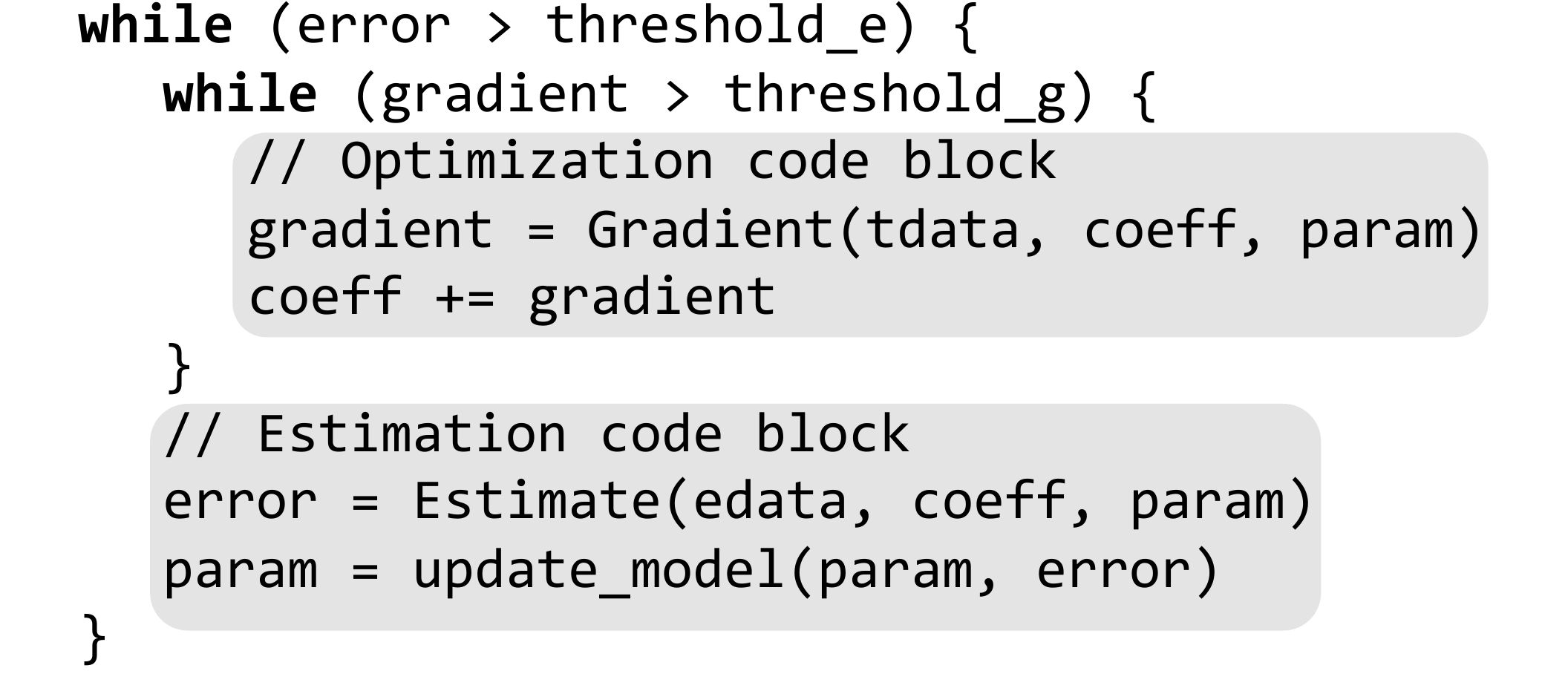}
\label{fig:driver-program}
}
&
\subfigure[Iterative execution graph.]
{
\includegraphics[width=3.0in]{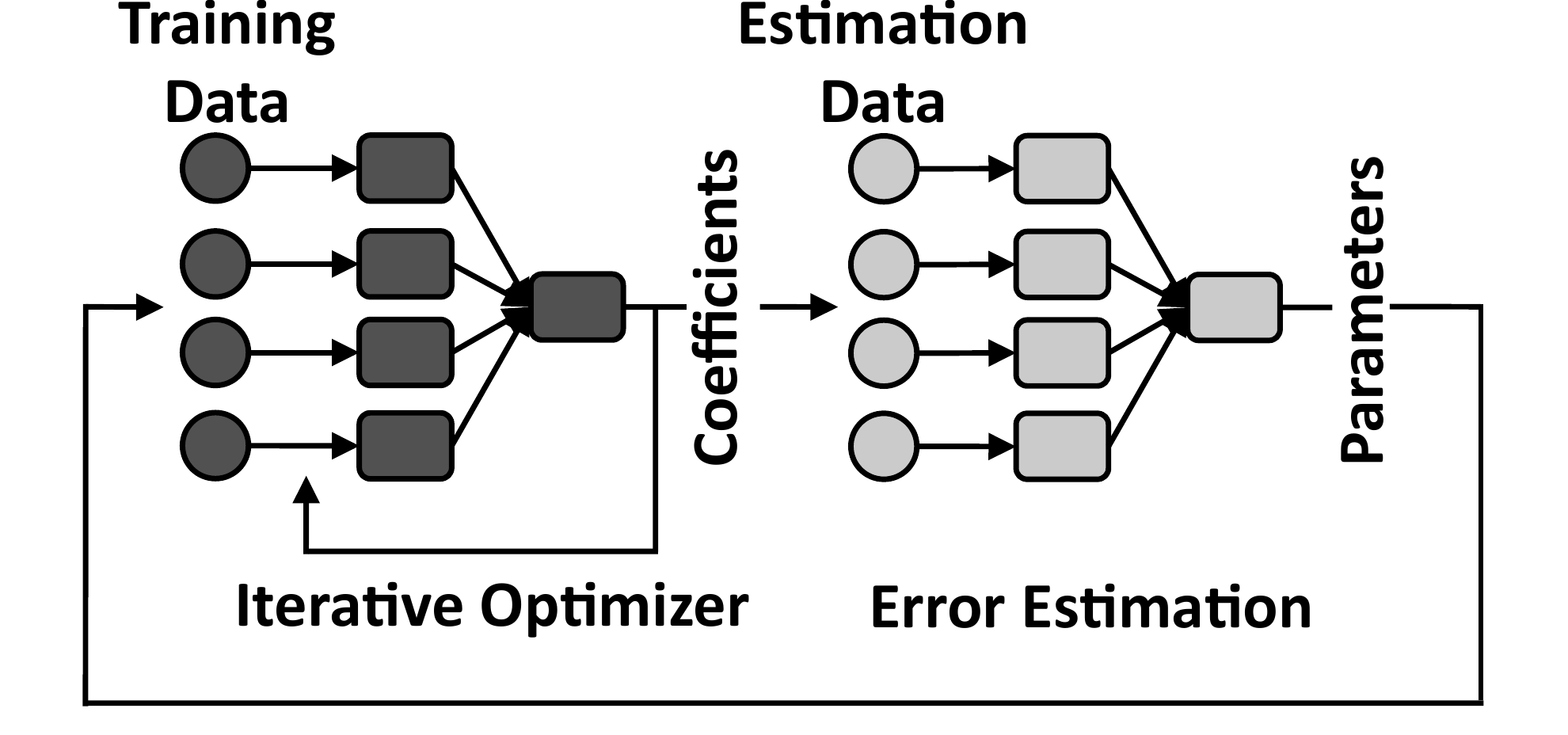}
\label{fig:task-graph}
}
\end{tabular}

\caption{Task graph and driver program pseudocode of a training
  regression algorithm. It is iterative, with an outer loop for
  updating model parameters based on the estimation error, and an
  inner loop for optimizing the feature coefficients. The driver program
  has two basic blocks corresponding to inner and outer loops.  {\tt
    Gradient} and {\tt Estimate} are both parallel operations that
  execute many tasks on partitions of data.}
\label{fig:cross-validation}

\end{figure*}

This section introduces execution templates and their characteristics.
Figure~\ref{fig:architecture} shows the general architecture of cloud computing
systems. Execution templates operate on the controller and its interfaces.

The template abstraction is motivated by the fact that long-running jobs are
usually iterative and run same set of tasks repetitively~\cite{ernest} with
minor changes. For example, Figure~\ref{fig:cross-validation} shows the
pseudocode and task graph for a training regression algorithm. The algorithm
consists of a nested loop.  The {\tt Gradient} and {\tt Estimate} operations
can each generate many thousands of tasks. This graph structure is identical
for each iteration, but the same vertex in two iterations can have different
values across iterations, such as the {\tt coeff} and {\tt param} parameters.
Furthermore, task identifiers change across iterations.
With execution templates, the control plane can
leverage the fixed structure to improve the performance.

\subsection{Abstraction}\label{sec:abstraction}

An Execution template is a parameterizable list of tasks. The fixed structure
of the template includes the list of tasks, their executable functions, task
dependencies, relative ordering, and data access references. The parameter list
includes the task identifiers and runtime parameters passed to each task.

The number of tasks listed in an execution template affects
the control flow flexibility. To enable data dependent branches and nested loop
structures, execution templates work at the granularity of basic blocks. A {\it
basic block} is a code sequence in the driver program with only one entry point
and no branches except the exit. For example, Figure~\ref{fig:cross-validation}
has two basic blocks, one for the inner loop and one for the outer loop
operations. Note that loop unrolling and other batching techniques~\cite{drizzle}
cannot capture nested loops and data dependent branches.

Execution templates are {\it installed} and {\it instantiated} at run time.
These two operations results in performance improvements in the control plane
by caching and reusing repetitive control flow.  Execution templates also
support two special operations, {\it edits} and {\it patching}, which deal with
scheduling changes and dynamic control flow. Each operation is discussed in the
following subsections.

\subsection{Installation and Instantiation}

There are two types of execution templates, one for the driver-controller
interface called a {\it controller template}, and one for the controller-worker
interface called a {\it worker template}. Controller templates contain the
complete list of tasks in a basic block across all of the worker nodes.  They
cache the results of creating tasks, dependency analysis, data lineage,
bookkeeping for fault recovery, and assigning data partitions as task
arguments. For every unique basic block, a driver program installs a controller
template at the controller. The driver can then execute the same basic block again
by telling the controller to instantiate the template.

Where controller templates describes a basic block over the whole system, each
worker template describes the portion of the basic block that runs on a
particular worker. Workers cache the dependency information needed for a worker
to execute the tasks and schedule them in the right order. Like
TensorFlow~\cite{tensorflow}, external dependencies such as data exchanges,
reductions, or shuffles appear as tasks that complete when all data is
transferred. Worker templates include metadata identifying where needed data
objects in the system reside, so workers can directly exchange data and execute
blocks of tasks without expensive controller lookups.

When a driver program instantiates a controller template, the controller makes
a copy of the template and fills in all of the passed parameters. It then
checks whether the prior assignment of tasks to workers matches existing worker
templates. If so, it instantiates those templates on workers, passing the
needed parameters.  If the assignment has changed, it either edits worker
templates or installs new ones.  In the steady state, when two iterations of a
basic block run on the same set of $n$ workers, the control plane sends $n+1$
messages: one from the driver to the controller and 1 from the controller to
each of the $n$ workers.

\subsection{Edits}

Execution templates have two mechanisms to make control plane
overhead scale gracefully with the size of scheduling changes:
installing new templates and editing existing ones. 
If the controller makes large changes to a worker's tasks, it can
install a new worker template.  Workers cache multiple worker
templates, so a controller can move between several different
schedules by invoking different sets of worker templates.

\begin{figure}
\centering
\setlength\tabcolsep{0pt}
\begin{tabular}{cc}
\subfigure[Edit migrating task from 2 to 1]
{
\includegraphics[height=1.8in]{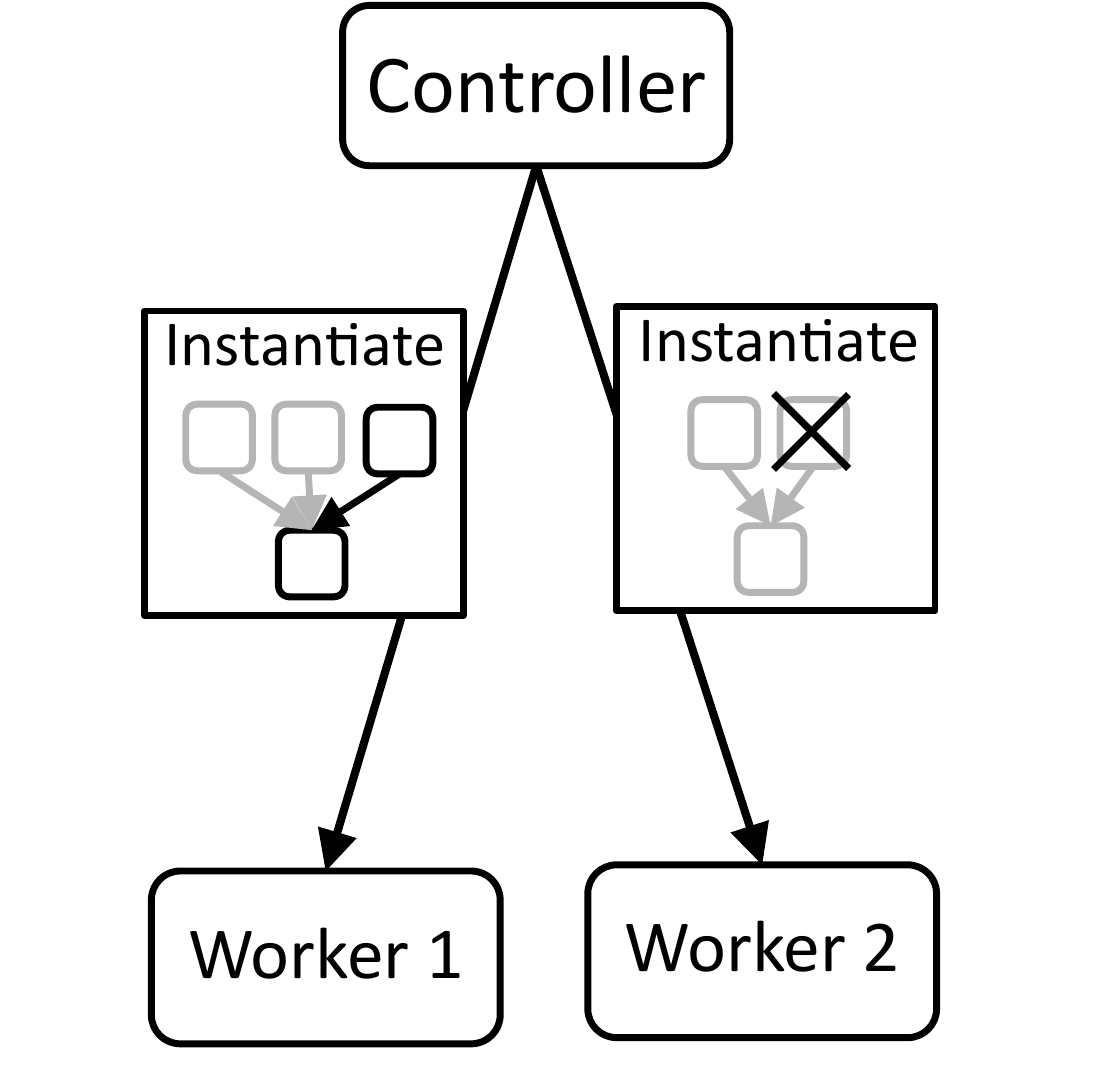}
\label{fig:edit}
}
&
\subfigure[Patching]
{
\includegraphics[height=1.8in]{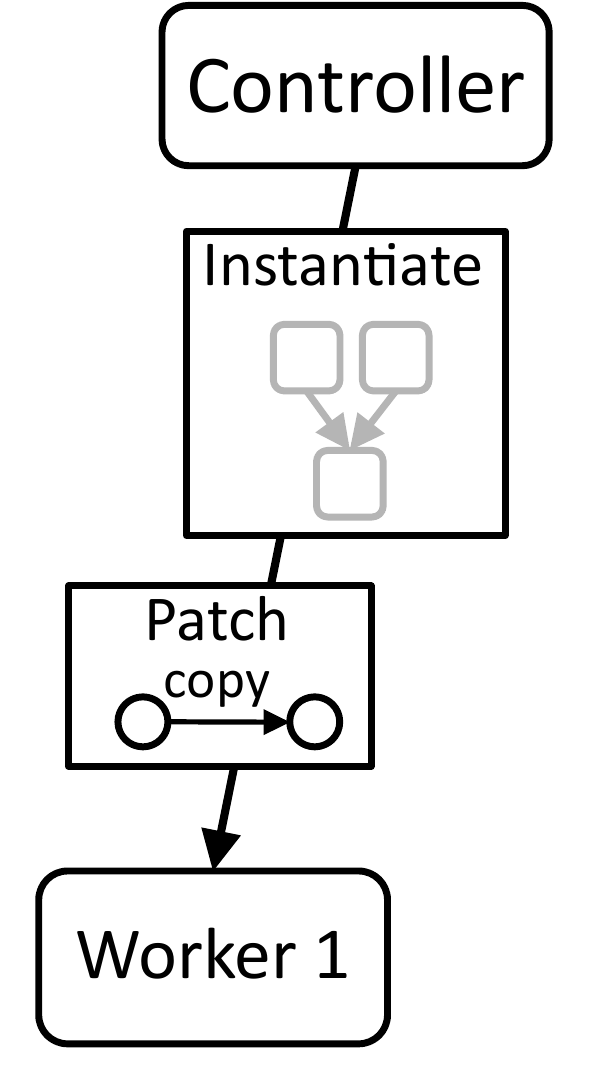}
\label{fig:patch}
}
\end{tabular}
\caption{Patches and edits allow a framework to efficiently adapt
  templates to dynamic changes in the system. Patches move and copy
  data objects to match a template's preconditions, while edits
  dynamically modify a template in place. Grey denotes cached template
  information, while black denotes information sent over the network.}
\label{fig:fine-grained}
\end{figure}

Edits allow a controller to change an existing worker template.
Figure~\ref{fig:edit} shows how edits manifest in the 
control plane: they modify already installed templates in place.
Edits are used when the controller needs to make small changes to the
schedule, e.g., migrate one of many partitions.  Edits are included as
metadata in a worker template instantiation message and modify its data
structures. An edit can remove and add tasks. Edits keep the cost of dynamic
scheduling proportional to the extent of changes. 
If large changes are needed, the controller can install new templates.

\subsection{Patching}

Installation and edits allow a controller to make fine-grained
changes to how a basic block is distributed across workers.
Patches allow templates to efficiently handle
dynamic program control flow. This is important when loop
conditions are based on data, such as running until an error value falls
below a threshold.

Each worker template has a set of preconditions that must hold when the
template is instantiated, for example requiring certain data objects to reside
in local memory.  When a driver program instantiates a controller template, the
system state may not meet the preconditions of the associated worker templates.
Since the driver program controls job execution and decides which templates to
execute next, the controller has to react to the driver's stream of controller
template instantiation requests and enforce the preconditions on the fly.

Controller uses patching to update and move data from one worker to another to
satisfy the preconditions. For example, the worker templates for the inner loop
in Figure~\ref{fig:driver-program} have the
precondition that {\tt param} needs to be in local memory.
But there are two cases in which the controller might invoke
the templates: the first iteration of the loop and subsequent 
iterations. In subsequent iterations, {\tt param} is inductively
already in local memory.  However, on the first iteration, {\tt
  param} exists only on the worker that calculated it. The controller
therefore {\it patches} the inner loop template, sending directives to
workers that copy {\tt param} to each worker (Figure~\ref{fig:patch}).

Patching is necessary because a basic block can be entered from many
different positions in code. When a template is created, the
controller may not even have seen all of these positions (e.g., an
edge case covered by an if/else). There are two options to deal with
uncertainties in control flow. The controller can either ensure that
the preconditions of every template always hold, or when a template is
instantiated it can patch system state to match the preconditions. The
first approach is prohibitively expensive, because it requires
unnecessary and expensive data copies. E.g., it would require
immediately copying {\tt param} in Figure~\ref{fig:driver-program} to
every worker after it is calculated even if the outer loop terminates.

\section{System Design}
\label{sec:design}

This section defines the requirements that execution templates place on
a control plane and describes the design of a cloud computing framework,
called \name, that meets these requirements.

\subsection{Control Plane Requirements}\label{sec:requirements}

Conceptually, execution templates can be incorporated into any
existing cloud framework. Incorporating them, however, assumes certain
properties in the framework's control plane. We describe these
requirements here, and defer a discussion of how they can be
incorporated into existing systems to Section~\ref{sec:discuss}.

\medskip\noindent
1. Workers maintain a queue of tasks and locally determine when tasks are
runnable.  Worker templates create many tasks on a worker, most of which are
not immediately runnable because they depend on the output of prior tasks. A
worker must be able to determine when these tasks are runnable without going
through a central controller, which would become a bottleneck.

\medskip\noindent
2. Workers can directly exchange data. Within a single template, one worker's
output can be the input of tasks on other workers. As part of executing the
template, the two workers need to exchange data without going through a
central controller, which would become a bottleneck.

\medskip\noindent
3. Controller schedules fine-grained tasks. Fine-grained tasks are a
prerequisite to support fine-grained scheduling; they define the minimum
scheduling change that a system can support.

%\begin{table}
%{\scriptsize
%\begin{tabular}{lccc} \toprule[2pt]
%{\bf System} & {\bf Worker queue} & {\bf Worker exchange} & {\bf Fine-grained} \\ \midrule[1pt]
%Spark      &                  &                   & X \\ 
%Naiad      & X                & X                 & \\
%TensorFlow & X                & X                 & \\
%\name     & X                & X                 & X \\ \bottomrule[2pt]
%\end{tabular}
%}
%\caption{Spark, Naiad, TensorFlow, and other existing systems do not
%simultaneously meet all three control plane requirements. \name meets
%all three.}
%\label{tab:requirements}
%\end{table}

\subsection{\name Architecture}

%Table~\ref{tab:requirements} shows how Naiad, Spark, and TensorFlow
%meet the requirements that execution templates place on the control
%plane. No system meets all three. Section~\ref{sec:discuss} discusses
%some potential implications of the necessary changes to existing
%systems.

This section describes the design of \name, an analytics framework
that meets all three requirements. \name is 30,000 semicolons of C++
code and supports tasks written in C++. \name's system architecture
is designed to support execution templates. Like Spark, Naiad, and
TensorFlow, \name is designed to run computationally intensive jobs
that operate on in-memory data across many nodes.  Like Spark, \name
has a centralized controller node that receives tasks from a driver
program. The controller dispatches these application tasks to workers.
The controller is responsible for transforming tasks from a driver
program into an execution plan, deciding on which workers to run which
computations.

As it sends application tasks to workers, the controller inserts additional
control tasks, such as tasks to copy data from one worker to another. These
tasks explicitly name the workers involved in the transfer, such that workers
can directly exchange data.

\subsection{\name Data Model}

\name has an execution  model similar to Spark~\cite{spark}. A job is decomposed into
{\it stages}. Each stage is a computation over a set of input data and
produces a set of output data. Each data set is partitioned into many
{\it data objects} so that stages can be parallelized. Each stage
typically executes as many {\it tasks}, one per object, that operate
in parallel. In addition to the identifiers specifying the data
objects it accesses, each task can be passed parameters, such as a
model parameter or constants.

\name tasks operate on mutable data objects. Supporting in-place
modification of data avoids data copies and are crucial for
computational efficiency because writes and reads operate
on the same cache line. In-place
modification also has two crucial benefits for execution templates.
First, multiple iterations of a loop access the same objects and reuse
their identifiers. This means the data object identifiers can be
cached in a template, rather than be a run-time parameter.  This makes
templates more efficient to parameterize, as the object identifiers
can be cached rather than recomputed on each iteration. Second,
mutable data objects reduce the overall number of objects in the
system by a large constant factor, which improves lookup speeds.

Mutable objects mean there can be multiple copies and versions of an
object in the system. For example, for the code in
Figure~\ref{fig:driver-program}, after the execution of the outer
loop, there are $n$ copies of {\tt param}, one on each worker.
However, one copy of {\tt param}, has been written to, and has an updated
value. Each data object in the system therefore combines an object
identifier with a version number. The \name controller ensures,
through data copies, that tasks on a worker always read the latest
value according to the program's control flow.

\subsection{\name Control Plane}

The \name control plane has four major commands. Data commands
create and destroy data objects on workers. Copy commands copy data
from one data object to another (either locally or over a network).
File commands load and save data objects from durable storage.
Finally, task commands tell the worker to execute an application
function.

Commands have five fields: a unique identifier, a {\it read set} of
data objects to read, a {\it write set} of data objects to write, a
{\it before set} of the commands that must complete before this one
can execute, and a binary blob of {\it parameters}. Task commands
include a sixth field, which application function to execute.

A command's before set includes only other tasks on that worker. If
there is a dependency on a remote command, this is encoded through a
copy command. For example a task associated with the {\tt
  update\_model} operation in Figure~\ref{fig:driver-program} depends
on the results of the parallel {\tt Estimate} operation. The {\tt
  update\_model} task has $n$ copy commands in its before set; one for each
locally computed {\tt error} in each partition.

Copy commands execute asynchronously and follow a push model. A sender
starts transmitting an object as soon as the command's before set is
satisfied. Because this uses asynchronous I/O it does not block a
worker thread. Similarly, a worker asynchronously reads data into
buffers as soon as it arrives. Once the before set of a data receive
job is satisfied (the new object is safely visible to the worker), it
changes a pointer in the data object to point to the new buffer.

\section{Implementation}
\label{sec:implementation}

\begin{figure*} 
\centering
\setlength\tabcolsep{-3pt}
\begin{tabular}{cc}
\subfigure[ A controller template represents the common structure of a task
graph metadata. It stores task dependencies and data access patterns. It is
invoked by filling in task identifiers and parameters to each task.]
{
\includegraphics[width=3.0in]{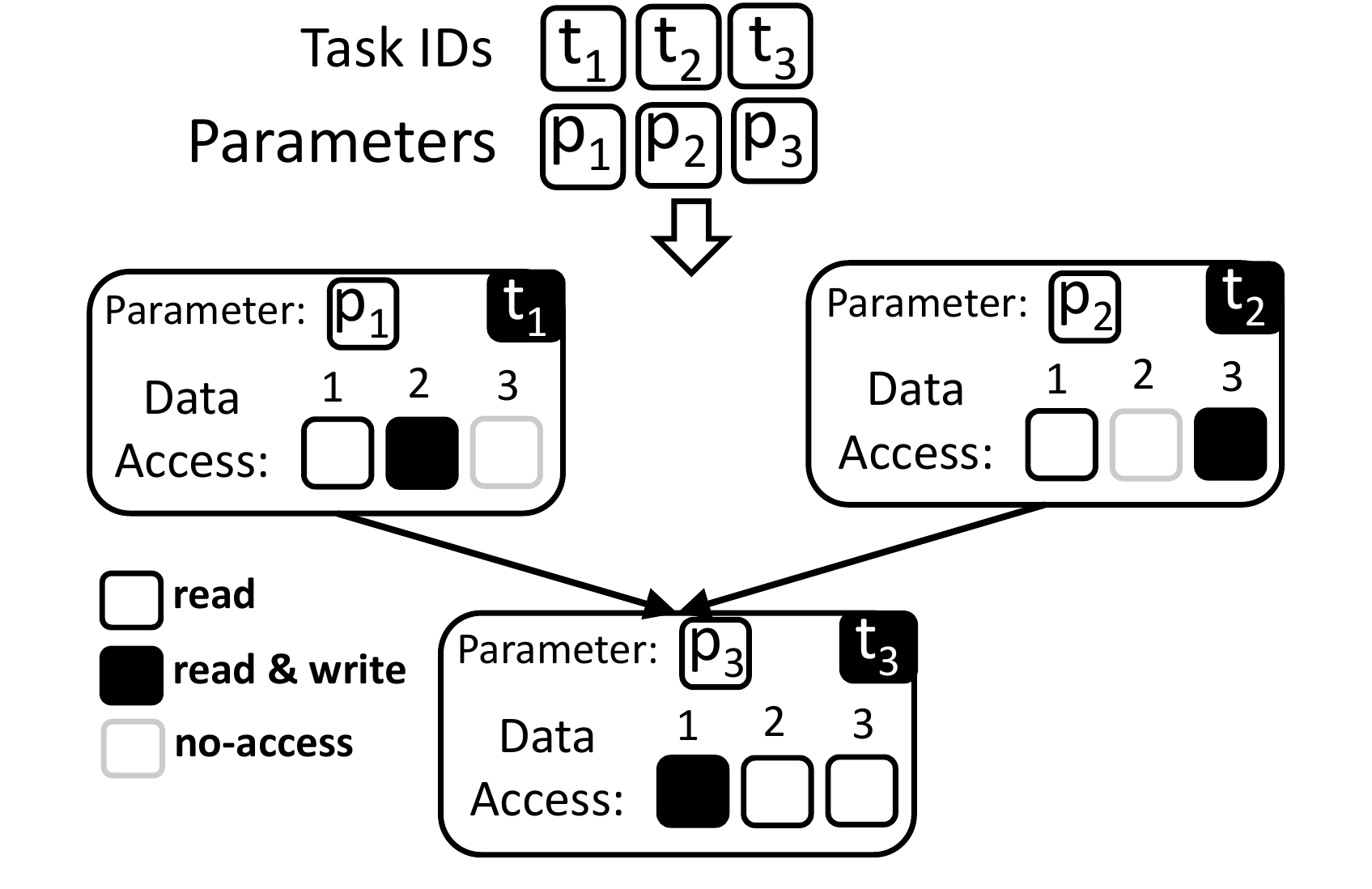}
\label{fig:controller-template}
}
&
\subfigure[Each worker template stores the common structure of a task graph for
execution including the data copies among workers. It is invoked by passing the
task identifiers, and parameters to each task.]
{
\includegraphics[width=3.0in]{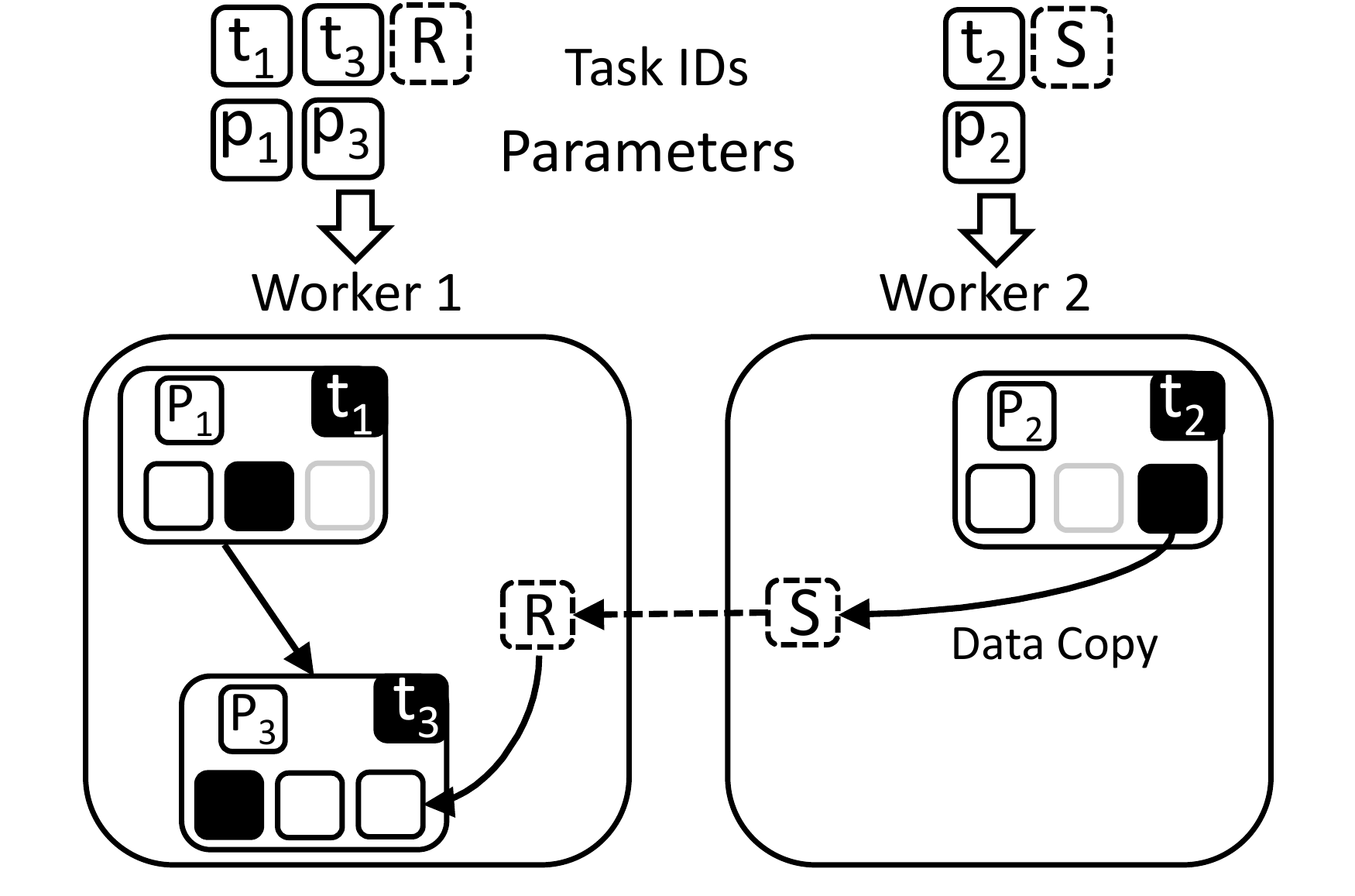}
\label{fig:worker-template}
}
\end{tabular}
\caption{Controller template (a) and worker templates (b) for a simple task graph.}
\end{figure*}

% This is hoe I counted the semicolons:
% cd cloud/src/nimbus/src
% grep ";" -r * --exclude-dir="obsolete" --exclude-dir="data" | grep -v "//" | wc -l

This section describes how \name implements execution templates and
their operations.  

\subsection{Installation and Instantiation}\label{sec:instantiation}

Template installation begins with the driver sending a start template message
to the controller at the beginning of a basic block.  In current implementation
of Nimbus, programmer explicitly marks the basic block in the driver program;
one can imagine other automatic approaches such as static program analysis.  As
the controller receives tasks, it simultaneously schedules them normally and
stores them in a temporary task graph data structure.

At the end of the basic block, the driver sends a template finish message. On
receiving a finish message, the controller takes the task graph and
post-processes it into an optimized, table-based data structure. Pointers are
turned into indexes for fast lookups into arrays of values.

Controller templates cache the read set, write set, and function identifier. A
template instantiation message includes an array of command identifiers and a
block of task parameters. Within a template, task identifiers index into this
array.  The one time cost of generating the ordered indices keeps the
successive instantiations efficient.  Figure~\ref{fig:controller-template}
shows the instantiation of a controller template with new set of task
identifiers and parameters.

Once it has generated the controller template, the controller generates the
associated worker templates. Worker templates have two halves. The first half
exists at the controller and represents the entire execution across all of the
workers.  This centralized half allows the controller to cache how the
template's tasks are distributed across workers and track the preconditions for
generating patches when needed.

Worker template has a preconditions list which data objects at each worker must
hold the latest update to that object. An important detail is that not all data
objects are required to be up to date: a data object might be used for writing
intermediate data and be updated within the worker template itself. For
example, in Figure~\ref{fig:worker-template}, the third data object on worker 1
does not need to have the latest update at the beginning of the worker
template; the data copy within the worker template updates it.

The second half of the worker template is distributed across the workers and
caches the per-worker local command graph which they locally schedule. The
controller installs worker templates very similarly to how the driver installs
controller templates. And like controller templates, instantiation passes an
array of task identifiers and parameters.  Figure~\ref{fig:worker-template}
shows a set of worker templates for controller template in
Figure~\ref{fig:controller-template}.

\subsection{Patching}\label{sec:patching}

Before instantiating a worker template, controller must {\it validate} whether
the template's preconditions hold and patch the worker's state if not.
Validating and patching must be fast, because they are sequential control plane
overhead that cannot be parallelized. Making them fast is challenging, however,
when there are many workers, data objects, and tasks, because they require
checking a great deal of state.

\name uses two optimizations to keep validation and patching fast.
The first optimization relates to template generation. When generating
a worker template, \name ensures that the precondition of the
template holds when it finishes. By doing so, it ensures that tight
inner loops, which dominate execution time and control plane traffic,
automatically validate and need no patching. As an example, in
Figure~\ref{fig:worker-template}, this adds a data copy of object 1 to worker
2 at the end of the template.

Second, workers cache patches and the controller can invoke these
patches much like a template. When a worker template fails validation,
the controller checks a lookup table of prior patches indexed by what
executed before that template. If the cached patch will correctly patch
 the template, it sends a single command to the worker to instantiate
the patch. Otherwise, it calculates a new patch and sends all of the
resulting commands. We have found that the patch cache has a very high
hit rate in practice because control flow, while dynamic, is typically
quite narrow.

\begin{figure}
\centering
\setlength\tabcolsep{0pt}
\begin{tabular}{cc}
\subfigure[Before.]
{
\includegraphics[height=1.5in]{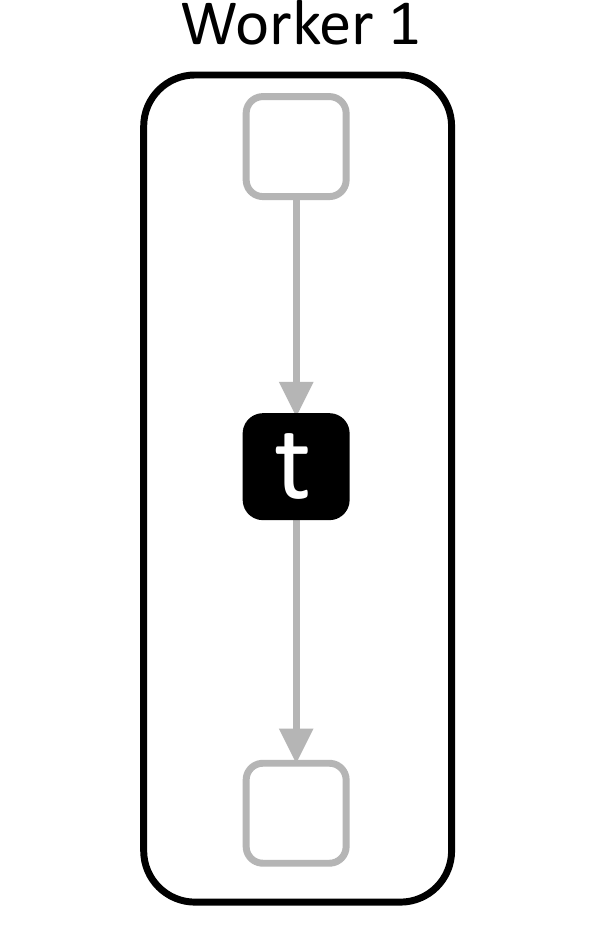}
\label{fig:edit-before}
}
&
\subfigure[After.]
{
\includegraphics[height=1.5in]{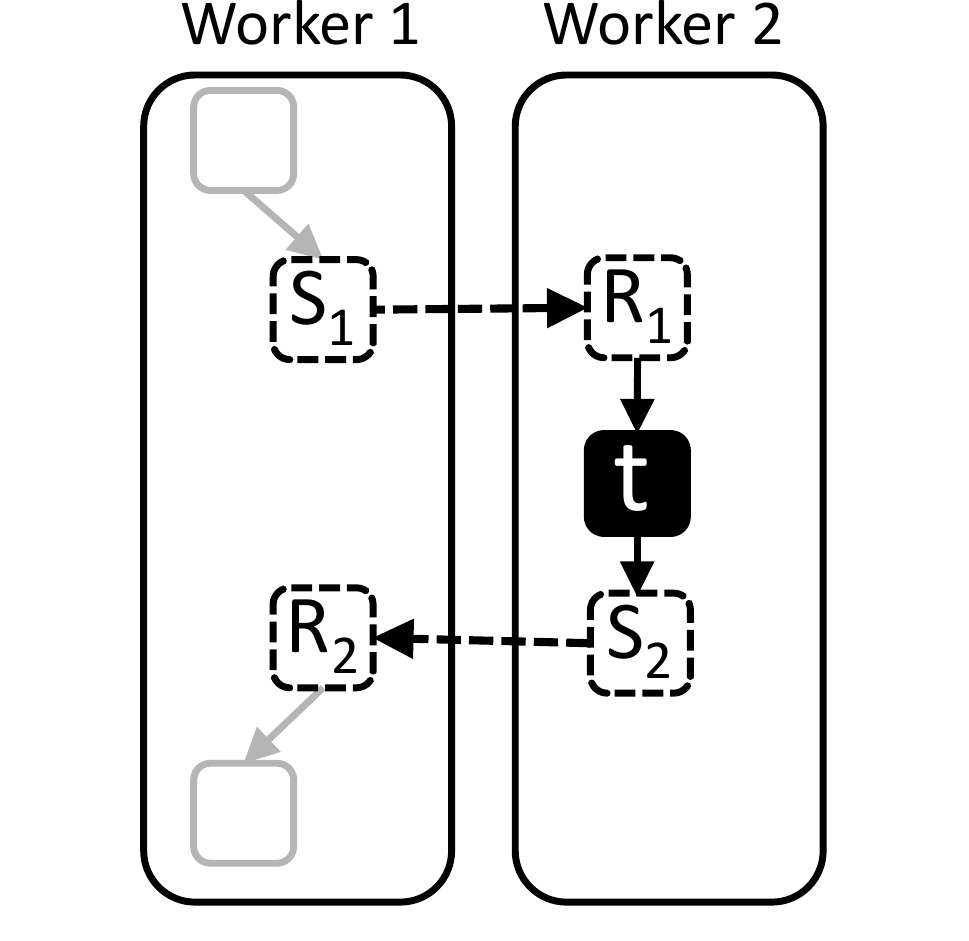}
\label{fig:edit-after}
}
\end{tabular}

\caption{Edits to migrate a task. The controller removes the task from
  worker 1's template and adds two data copy commands (S$_1$, R$_2$).
  It adds the task and two data copy commands (R$_1$, S$_2$) to worker
  2's template.}
\label{fig:edit-example}

\end{figure}

\subsection{Edits}\label{sec:edits}

Whenever a controller instantiates a worker template, it can attach a
list of edits for that template to apply before instantiation. Each
edit specifies either a new task to include or a task to remove.
Edits are usually limited to the actual tasks being added or removed,
because in cases when there are dependencies with other tasks, tasks
are exchanged with data copy commands. Figure~\ref{fig:edit-example}
shows, for example, how a task's entry in a before set is replaced by
a data receive command. As long as the data receive command is assigned
the same index within the command identifier array, other commands
do not need to change.
Using edits, minor changes in scheduling have very small costs and the
cost scales with the size of the change.

\subsection{Fault Recovery}

\name implements a checkpoint recovery mechanism.  Although a controller keeps
the full lineage for every data object in the system, for iterative
computations we found that linage-based recovery~\cite{spark} is essentially
identical to checkpointing because there are frequent synchronization points
around shared global values.  Any lineage recovery beyond a synchronization
point requires regeneration of every data object, which is a checkpoint.

\name automatically inserts checkpoints into the task stream from a driver
program. When a checkpoint triggers, the controller waits until all worker task
queues drain, stores a snapshot of the current execution graph, and requests
every worker to write its live data objects to durable storage.

When a controller determines a worker has failed (it stops sending periodic
heart beat messages or workers depending on its data fall idle), it sends a
halt command to every worker. On receiving the command, workers terminate all
ongoing tasks, flush their queues, and respond back.  Then, the controller
sends commands to load the latest checkpoint into memory, reverts to the stored
execution graph snapshot, and restarts execution.

\section{Evaluation}
\label{sec:evaluation}

This section evaluates execution templates in \name, comparing them with
Spark's fine-grained centralized scheduler, Naiad's high-throughput distributed
data flow graphs \footnote{TensorFlow's control plane design is very similar to
Naiad's which results in very close performance and behaviour.}, and application-level MPI
messaging.  In summary, our findings show:

\begin{itemize}[leftmargin=*] % [nolistsep, leftmargin=*]
  \setlength{\itemsep}{.5pt}

\item Execution templates allow \name to schedule hundreds of thousands
of tasks per second, imposing a control overhead competitive with Naiad's
distributed data flow graphs.

\item Execution templates allow \name to schedule at task granularity,
providing a runtime flexibility and adaptivity equivalent to Spark's
centralized scheduler.

\item Execution templates are expressive enough to support complex,
high-performance  applications, such as a particle-levelset water
simulation with a triply nested, data dependent loop
and tasks as short as $100\mu$s.

\end{itemize}

\subsection{Methodology}

All experiments use Amazon EC2 compute-optimized instances since they are the
cheapest option for compute-bound workloads. Worker nodes use {\tt c3.2xlarge}
instances with 8 virtual cores and 15GB of RAM. Controllers run on a more
powerful {\tt c3.4xlarge} instance to show how jobs bottleneck on the
controller even when it has more resources. All nodes are allocated in a single
placement group and so have full bisection bandwidth.

We compare the performance of \name with Spark 2.0 and Naiad 0.4.2 using two
machine learning benchmarks, logistic regression and k-means clustering. We
measure iteration time and control plane overhead on 20-100 worker nodes.
Because our goal is to measure the task throughput and scheduling granularity
of the control plane, we factor out language differences between the three
frameworks and have them run tasks of equal duration.  We chose the task
duration as the fastest of the three frameworks, as it evaluates the highest
task throughput. \name tasks run 8 times faster than Spark's MLlib due to Spark
using a JVM (a 4x slowdown) and its immutable data requiring copies (a 2x
slowdown).  \name tasks run 3 times faster than Naiad due to Naiad's use of the
CLR.  To show that tasks in Naiad and Spark are not CLR or Scala codes but
rather tasks that run as fast as C++ ones, we label them {\it Naiad-opt} and
{\it Spark-opt}. This is done by replacing the task computations with a spin
wait as long as C++ tasks.
% This evaluates the performance of these frameworks
% if they called directly into native code with no overhead.

The Naiad and \name implementations of k-means and logistic regression include
application-level two-level reduction trees.
% Workers locally reduce their
% results before passing a single partial reduction to the global reducer.
Application-level reductions in Spark harm completion time because they add
more tasks that bottlenecks at the controller.

\begin{table}
\centering
{\small 
\begin{tabular}{lr} \toprule[2pt] 
                                            & {\bf Per-task cost} \\
\midrule[1pt]
Installing controller template              & $25\mu$s  \\
Installing worker template on controller    & $15\mu$s  \\
Installing worker template on worker        & $9\mu$s   \\
\midrule[1pt]
\name schedule task                        & $134\mu$s  \\
Spark schedule task                         & $166\mu$s  \\
\bottomrule[2pt]
\end{tabular}
}
\caption{Template installation is fast compared to scheduling.
  The 49$\mu$s per-task cost is evenly split between the controller and
  worker templates. 
  Installing a new worker template has a per-task cost
  of 24$\mu$s, and 18\% overhead on centrally scheduling that
  task.}
\label{tab:installation-costs}
\end{table}

\begin{table}
\centering
{\small 
\begin{tabular}{lr} \toprule[2pt] 
                                            & {\bf Per-task cost}         \\
\midrule[1pt]
Instantiate controller template             & $0.2\mu s$                  \\
Instantiate worker template                 & \multirow{2}{*}{$1.7\mu s$} \\
(auto-validation)                                                \\
Instantiate worker template                 & \multirow{2}{*}{$7.3\mu s$}  \\
(validation)                                                \\
\bottomrule[2pt]
\end{tabular}
}
\caption{Template instantiation is fast. For the common case of a template
automatically validating (repeated execution of a loop), instantiation takes
1.9$\mu$s/task: \name can schedule over 500,0000 tasks/sec. If dynamic control
flow requires a full validation, it takes $7.5\mu$s/task and \name can
schedule 130,000 tasks/second.}
\label{tab:instantiation-costs}
\end{table}

\begin{table}[t]
\centering
{\small 
\begin{tabular}{lr} \toprule[2pt] 
                                               & {\bf Cost} \\
\midrule[1pt]
Nimbus single edit                             & $\approx41\mu s$  \\
Nimbus 5\% task migration (800 edits)        & $35ms$     \\
Nimbus complete installation (8000 tasks)      & $203ms$    \\
\midrule[1pt]
Naiad any change                               & $230ms$    \\
\bottomrule[2pt]
\end{tabular}
}
\caption{A single edit to the logistic regression job takes 41$\mu$s Nimbus, and the
cost scales linearly with the number of edits. Edits are still less expensive
than full installation when migrating as high as 5\% of the template's tasks.
Any change in Naiad induces the full cost of data flow installation}
\label{tab:edit-costs}
\end{table}

\begin{figure*}
\centering
\setlength\tabcolsep{-3pt}
\begin{tabular}{cc}
\subfigure[Logistic regression]
{
\includegraphics[width=3in]{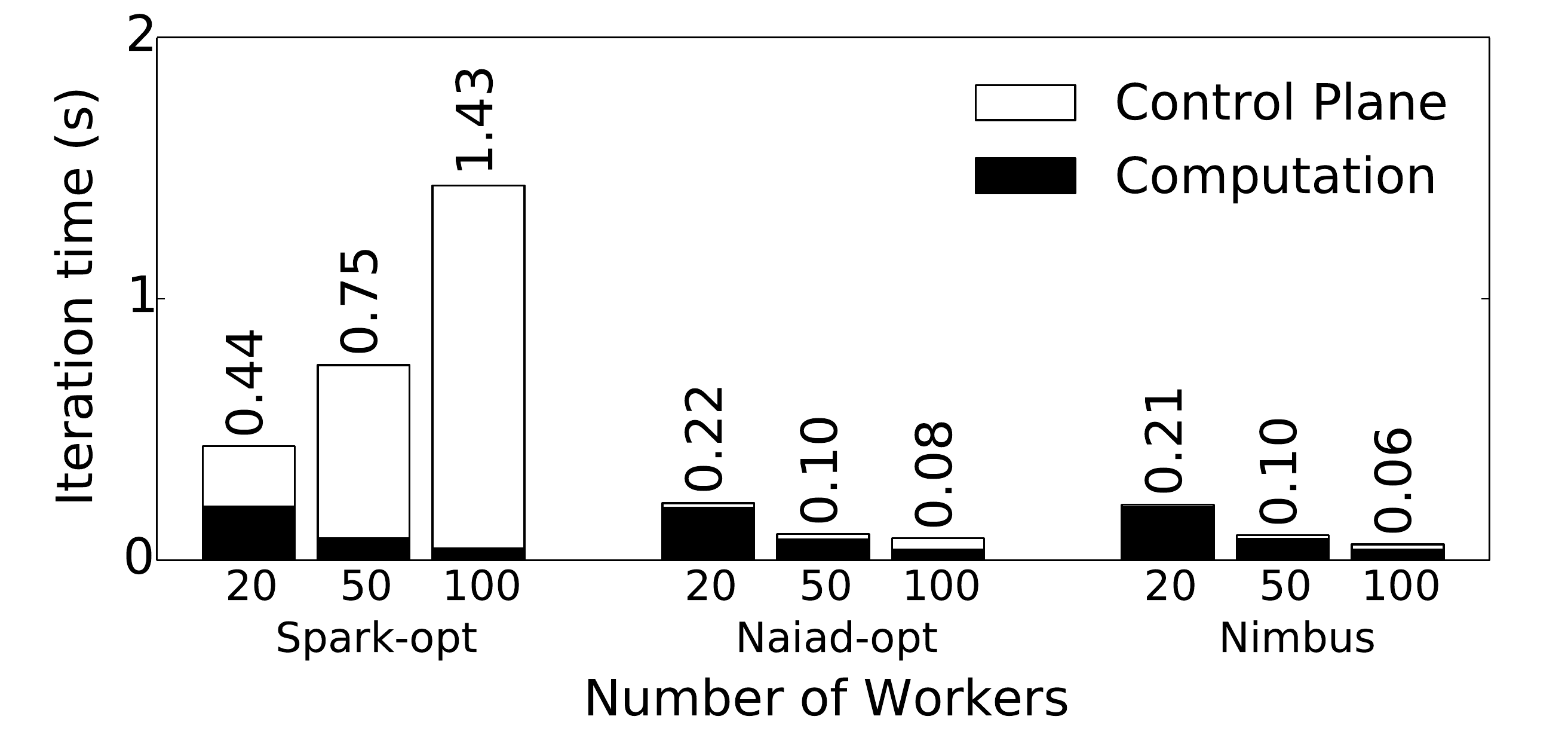}
\label{fig:lr-strong}
}
&
\subfigure[K-means clustering]
{
\includegraphics[width=3in]{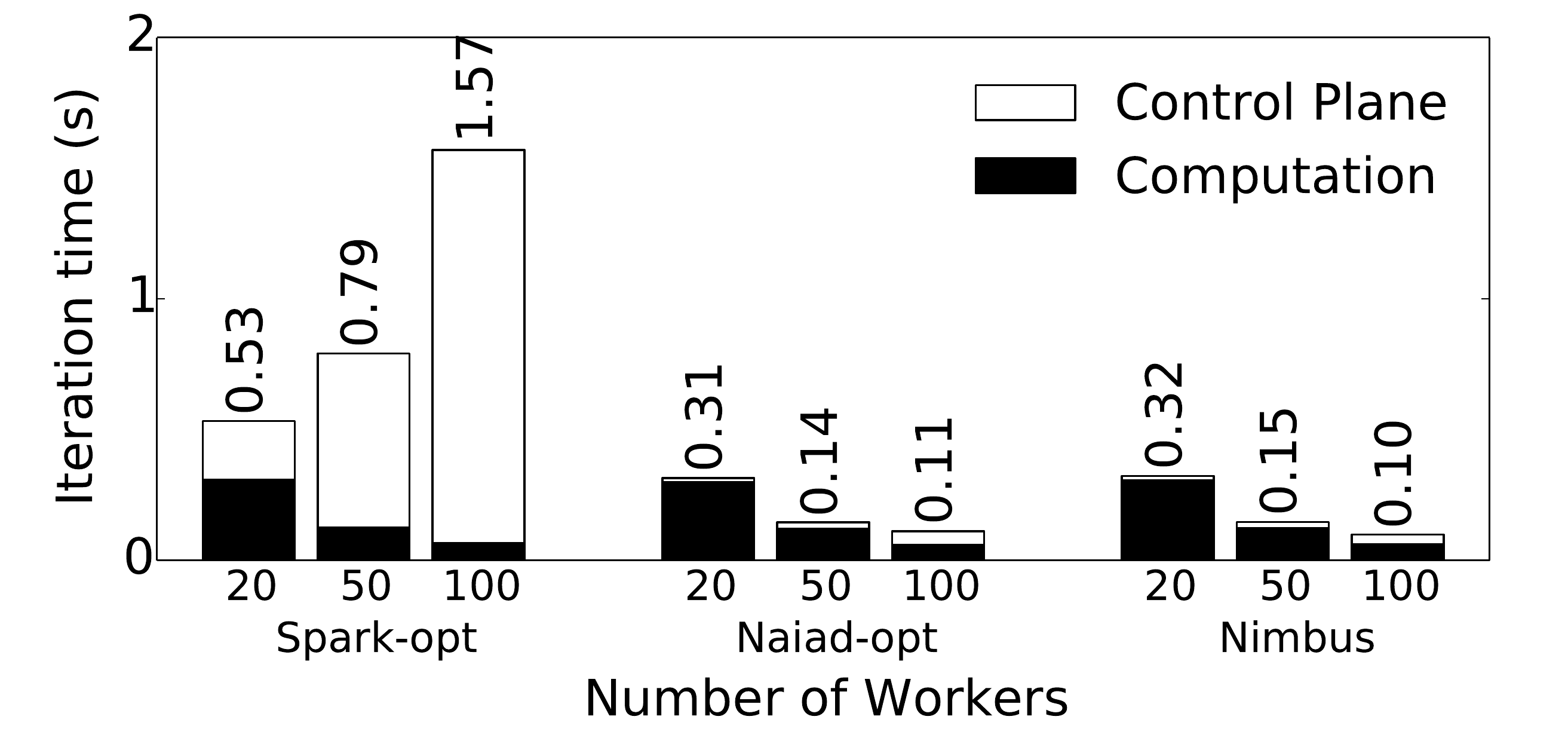}
\label{fig:kmeans-strong}
}
\end{tabular}
\caption{Iteration time of logistic regression and k-means for a data set of
size 100GB. \name executes tasks implemented in C++.  Spark-opt and Naiad-opt
show the performance when the computations are replaced with spin-wait as fast
as tasks in C++. Execution templates helps centralized controller of \name
scale out almost linearly.}
\label{fig:data-analytics}
\end{figure*}

\subsection{Micro-Benchmarks}
\label{sec:cost}

This section presents micro-benchmark performance results.  These results are
from a logistic regression job with a single controller template with 8,000
tasks, split into 100 worker templates with 80 tasks each.

Table~\ref{tab:installation-costs} shows the costs of template installation.
We report the per-task costs because they scale with the number of tasks (there
are individual task messages).  We also report the cost of centrally scheduling
a task in Spark and \name to give context.  Installing a template has a
one-time cost of installing the controller template and the potentially
repeated cost of installing worker templates. Adding a task to a controller
template takes $25\mu$s. Adding it to a worker template takes $24\mu$s. In
comparison to scheduling a task ($134\mu$s), this cost is small.  Installing
all templates has an overhead of 36\% on centrally scheduling tasks.

Table~\ref{tab:instantiation-costs} shows the costs of template instantiation.
There are two cases for the worker template. In the first (common) case, the
template validates automatically because it is instantiated after the same
template. Since \name ensures that a template, on completion, meets its
preconditions, in this case the controller can skip validation. In the second
case, a different  worker template is instantiated after the previous one, and
controller must fully validate the template. When executing the inner loop of a
computation, \name's scheduling throughput is over 500,000 tasks/second
(.2$\mu$s + 1.7$\mu$s per task).

% Instantiating a controller template is extremely cheap, 200ns per task. All
% this involves is copying the controller template structure and filling in the
% task identifiers and parameters. Instantiating a worker template that
% automatically validates takes 1.7$\mu$s per task, while with explicit
% validation it takes 7.3$\mu$s per task. When executing the inner loop of a
% computation, \name's scheduling throughput is over 500,000 tasks/second.

Table~\ref{tab:edit-costs} shows edit costs. A single edit (removing or adding
a task) takes 41$\mu$s\footnote{It is greater than the cost of installing a
task in a worker template (29$\mu$s) due to the necessary changes in the task
graph and inserting extra copy tasks (see Figure~\ref{fig:edit-example}).}.  Edits allow
controllers to inexpensively make small-scale changes to worker templates. For
example, 800 edits (e.g., migrating 5\% of the tasks) takes 67ms, fraction of
complete installation cost. The cost of installing physical graphs
on Naiad is about 230ms which would be induced for any changes.
 
% Edits allow controllers to inexpensively make small-scale changes to
% worker templates.
% Figure~\ref{fig:edit-costs} shows how the cost of edits compared to
% the cost of reinstalling templates. The cost of edits increases
% linearly with the number of edits involved. However, because the cost
% of an individual edit (41$\mu$s) is greater than the cost of
% installing a task in a worker template (29$\mu$s), when the change is
% large enough it is faster to install a new template.  Furthermore,
% since edits operate in place, they do not allow quickly switching
% between very different schedules (as in
% Figure~\ref{fig:multi-tenant}).  Heuristics on when to edit versus
% install new templates are an area of future work.

\subsection{Control Plane Performance}

This section evaluates the strong scalability of execution templates and it's
impact on job completion time.  Figure~\ref{fig:data-analytics} shows the
results of running logistic regression and k-means clustering over a
100GB input once data has been loaded and templates have been
installed.  We observed negligible variance in iteration times and
report the average of 30 iterations.

\name and Naiad have equivalent performance; with 20 workers, an
iteration of logistic regression takes 210-220ms and with 100 workers
it takes 60-80ms.  The slightly longer time for Naiad with 100 workers
(80ms) is due to the Naiad runtime issuing many callbacks for the
small data partitions; this is a minor performance issue and can be
ignored. For k-means clustering, an iteration across 20 nodes takes
310-320ms and an iteration across 100 nodes takes 100-110ms.
Completion time shrinks slower than the rate of increased parallelism
because reductions do not parallelize.

\begin{figure}
\centering
\includegraphics[width=3.0in]{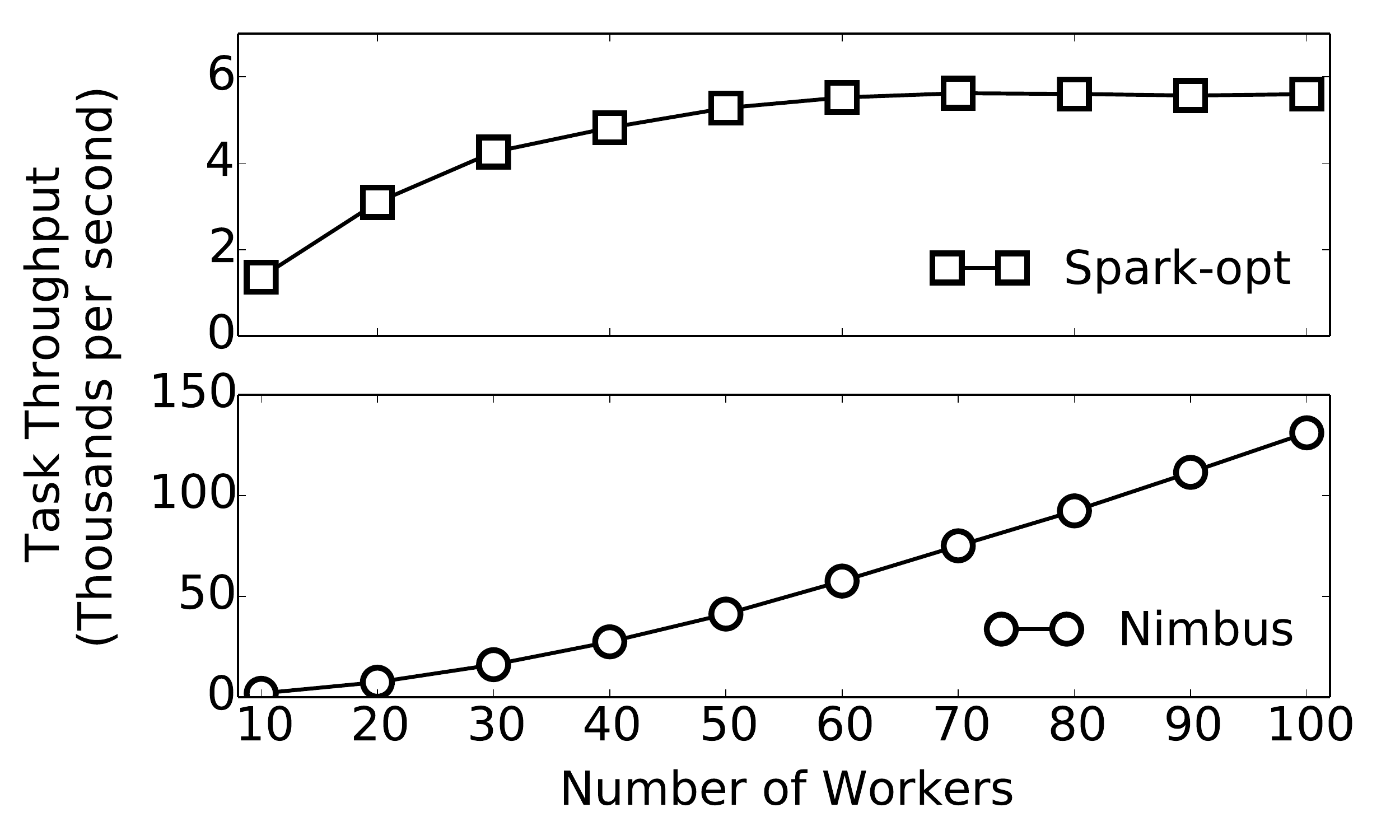}
\caption{Task throughput of \name and Spark as the number of workers
increases. Spark saturate at about 6,000 tasks per second, while \name grows
to adapt to the number of tasks required for more parallelism. Note that the
y-axis scale is different in the plots.}
\label{fig:task-throughput}
\end{figure}

Running over 20 workers, Spark's completion time is 70-100\% longer
than \name and Naiad. With greater parallelism (more workers), the
performance difference increases: Naiad and \name run proportionally
faster and Spark runs slower. Over 100 workers, Spark's completion
time is 15-23 times longer than \name. The difference is entirely due
to the control plane.  Spark
workers spend most of the time idle, waiting for the Spark controller
to send them tasks. In contrast, \name and Naiad have workers locally
generate and schedule tasks and so do not bottleneck at the controller.

Figure~\ref{fig:task-throughput} shows the rate at which \name and Spark 
schedule logistic regression tasks as the number of workers increases. Spark quickly
bottlenecks at 6,000 tasks per second. 
\name scales to support the increasing task throughput: a single
iteration over 100 workers takes 60ms and executes 8,000
tasks, which is 128,000 tasks/second (25\% of \name's maximum throughput).
Note that greater parallelism increases
the task rate superlinearly because it simultaneously creates more tasks
and makes those tasks shorter.

\subsection{Dynamic Scheduling}
\label{sec:fine-grained}

% This section evaluates how well execution templates can support fine-grained
% scheduling, and keep the cost proportional the size of changes.

% \subsubsection{Installation, Validation and Patching}

\begin{figure*}
\centering
\includegraphics[width=6.0in]{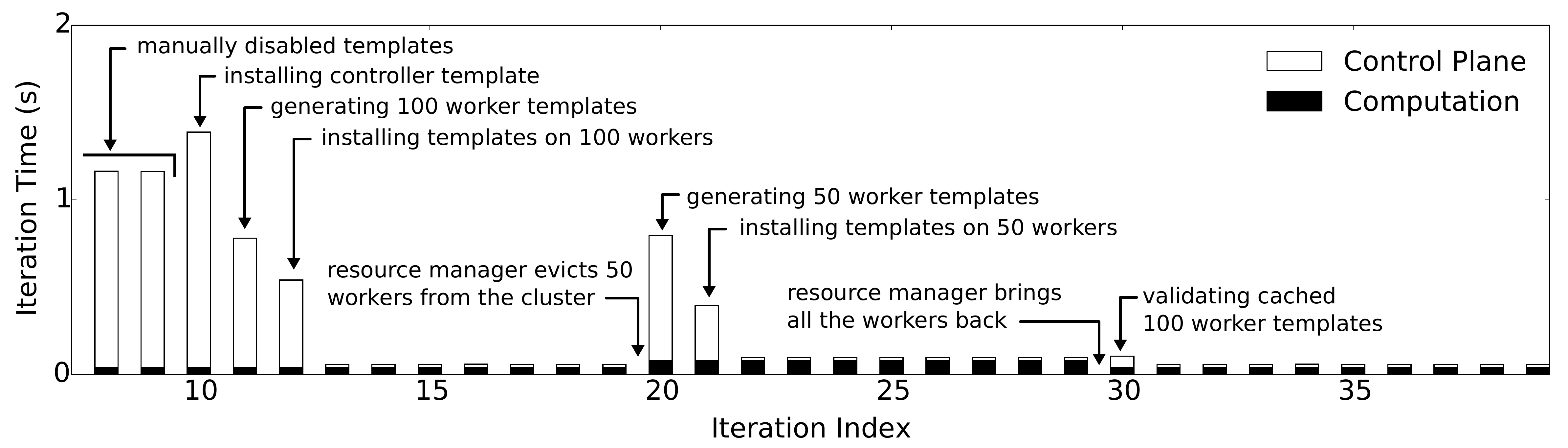}
\caption{Execution templates can schedule jobs with high task throughputs
while dynamically adapting as resources change. This experiment shows the
control overheads as a cluster resource manager allocates 100 nodes to
a job, revokes 50 of the nodes, then later returns them.}
\label{fig:multi-tenant}
\end{figure*}

Figure~\ref{fig:multi-tenant} shows the time per iteration of logistic
regression in \name as a cluster manager adjusts the available
resources.  The run starts with templates disabled: the control plane
overhead of a centralized scheduler dominates iteration time: each
iteration takes 1.07s.  At iteration 10, the driver starts using
templates. Iteration 10 takes $\approx1.3$s, as installing each of the
8,000 tasks in the controller template adds 25$\mu$s
(Table~\ref{tab:instantiation-costs}). On iteration 11, the controller
template has been installed, and the controller generates its half of
the worker template as it continues to send individual tasks to
workers.  This iteration is faster because the control traffic between
the driver and controller is a single instantiation message. On
iteration 12, the controller half of the worker templates has been
installed, and the controller sends tasks to and installs templates on
the workers.  On iteration 13, templates are fully installed and an
iteration takes 60ms (as in Figure~\ref{fig:lr-strong}), with minimal
control plane overhead.

At iteration 20, the cluster resource manager revokes 50 workers from
the job's allocation. On this iteration, the controller regenerates the
controller half of the worker template, migrating tasks from evicted
% workers to other workers that have their task's input data. On
workers to remaining workers. On
iteration 21, the controller generates new worker tasks for the 50 workers.
Execution time doubles because each worker is performing twice the work.
% To disambiguate the control plane and data migration overhead, the data
% sets on the evicted workers (50GB) are already loaded on the other half of
% workers (evenly among them).

At iteration 30, the cluster resource manager restores the 50 workers
to the job's allocation. The controller reverts to using the original
worker templates and so does not need to install templates. However,
on this first iteration, it needs to validate the templates. After
this explicit validation, each iterations takes 60ms.

% \subsubsection{Edits}

\begin{figure}
\centering
\includegraphics[width=3.0in]{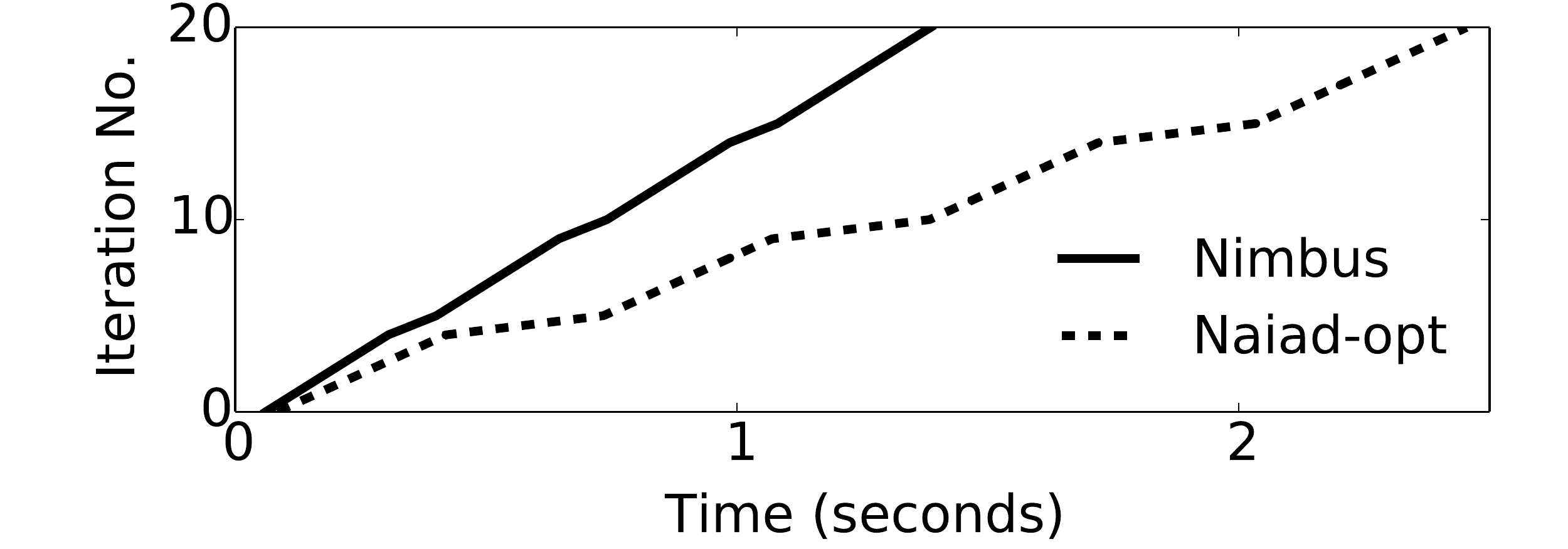}
\caption{Logistic regression over 100 workers with task migration every 5
iterations. Nimbus shows negligible overhead by using edits, while Naiad
requires complete data flow installation for migrations.}
\label{fig:dynamic-scheduling}
\end{figure}

Figure~\ref{fig:dynamic-scheduling} shows the scenario of running a logistic
regression job over 100 workers and migrating 5\% of tasks every 5 iteration.
Nimbus migration overhead is negligible, while Naiad requires complete
installation for any change in scheduling. Note that, current Naiad
implementation does not support any data flow flexibility once the job starts,
so the curve here is simulated from the numbers in Table~\ref{tab:edit-costs}
and Figure~\ref{fig:lr-strong}. The incremental edit cost lets Nimbus finish 20
iterations almost twice as fast as Naiad.

\subsection{Complex Applications}\label{sec:physbam}

To evaluate if execution templates can handle full applications with
complex control flows and data flows, we
use PhysBAM, an open-source computer graphics simulation
library~\cite{physbam}. It is the result of over 50 developer-years of
work and has won two Academy Awards. We ported PhysBAM to \name,
wrapping PhysBAM functions inside tasks and interfacing PhysBAM data
objects (level sets, mark-and-cell grids, particles) into \name. 

We ran a canonical particle-levelset fluid simulation benchmark, water
being poured into a glass~\cite{particle-levelset}. This is the same
core simulation used for the ocean in The Perfect Storm and the river
in Brave. It has a triply-nested loop with 21 different
computational stages that access over 40 different variables.  
Systems with static data flow (e.g., Naiad) cannot run this
simulation efficiently because the termination conditions 
of its two inner loops are based on data values. We ran a
$1024^{3}$ cell simulation (512GB-1TB of RAM) on 64 workers.  While
the majority of execution time is spent in tasks that take 60-70ms,
the median task length is 13ms, 10\% of tasks are $<$3ms and some
tasks are as short as 100$\mu$s.

\begin{figure}
\centering
\setlength\tabcolsep{0pt}
\setlength{\fboxrule}{0pt}
\begin{tabular}{c}
\subfigure[Still of water pouring in to a glass bowl.]
{
\fbox{\includegraphics[width=2.5in]{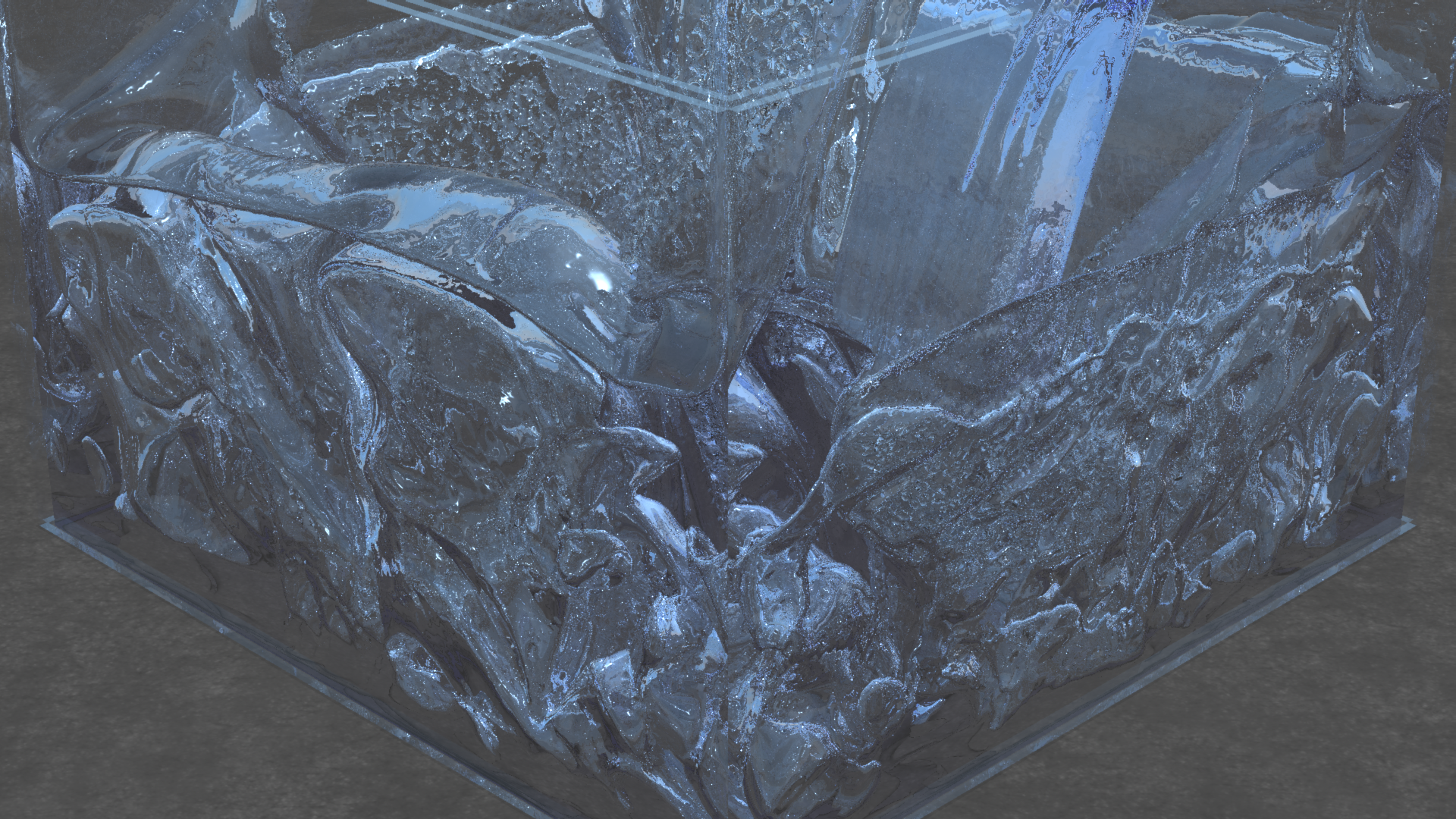}}
\label{fig:water}
}
\\
\subfigure[Iteration time of the main outer loop.]
{
\fbox{\includegraphics[width=3in]{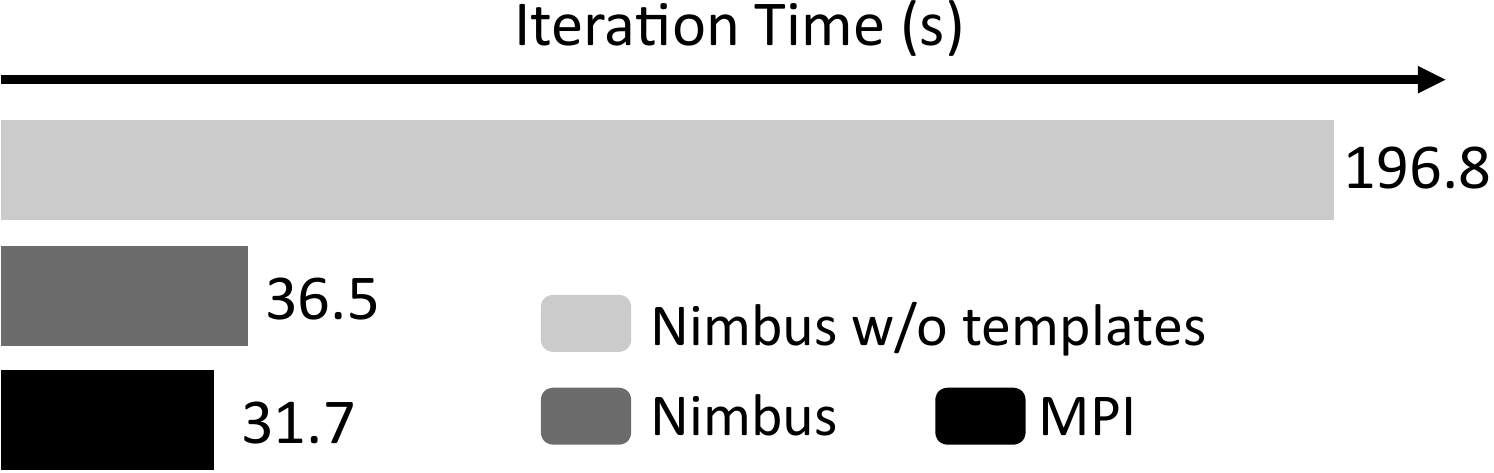}}
\label{fig:iteration}
}
\end{tabular}
\caption{PhysBAM water simulation.}
\label{fig:physbam}
\end{figure}

Figure~\ref{fig:physbam} shows the results of running the simulation
with PhysBAM's hand-tuned MPI libraries, in \name without templates
and in \name with templates. The MPI libraries cannot rebalance load, 
and in practice developers rarely use them due to their brittle behavior 
and lack of fault tolerance.  Without
templates, the central controller becomes the bottleneck and
the simulation takes 520\% longer than MPI.  With templates, the
simulation runs within 15\% of the MPI implementation, while providing
fine-grained scheduling, automatic fault tolerance, and adaptive load
balancing.

% \begin{figure}
% \centering
% \includegraphics[width=3in]{figs/physbam-code.pdf}
% \caption{Simplified Pseudo code of a canonical PhysBAM 
% water simulation in PhysBAM. The full particle-levelset algorithm 
% has 21 stages.}
% \label{fig:physbam-code}
% \end{figure}

% \begin{figure}
% \centering
% \includegraphics[width=3in]{figs/physbam-speedup.pdf}
% \caption{Iteration time of a PhysBAM water simulation.}
% \label{fig:physbam}
% \end{figure}

% \input{discussion}
\section{Related Work}
\label{sec:related}

We build on a large history of prior work that can be divided
into three major classes: cloud frameworks, cloud schedulers, and high
performance computing.

\medskip\noindent\textbf{Cloud frameworks} schedule tasks from a
single job.
% Cloud frameworks have a detailed understanding of the
% underlying data model, data placement, and performance trade-offs of
% individual tasks but can only reason about their single job.
Systems such as CIEL~\cite{ciel}, Spark~\cite{spark} and
Optimus~\cite{optimus} keep all execution state on a central
controller, dynamically dispatching tasks as workers become ready.
This gives the controller an accurate, global view of the job's
progress, allowing it to quickly respond to failures, changes in
available resources, and system performance. Execution templates
borrow this model, but cache scheduling decisions to drastically
increase scheduling throughput.

Systems such as Naiad~\cite{naiad} and TensorFlow~\cite{tensorflow}
take the opposite approach, statically installing an execution plan
on workers so the workers can locally generate tasks and directly
exchange data. Execution templates borrow this idea of installing
execution plans at runtime but generalize it to support multiple
active plans and dynamic control flow. Furthermore, execution
templates maintain fine-grained scheduling by allowing a controller to
edit the current execution plan.

Dataflow frameworks such as Dryad~\cite{dryad},
DryadLINQ~\cite{dryadlinq}, and FlumeJava~\cite{flumejava}, as well as
programming models such as DimWitted~\cite{dimwitted},
DMLL~\cite{dmll} and Spark optimizations~\cite{nvl-mit,
  nvl-platformlab,spark-sql, databricks-tungsten,databricks-spark2}
focus on abstractions for parallel computations that enable
optimizations and high performance, in some cases faster than
hand-written C.  This paper examines a different but complementary
question: how can a framework's runtime scale to support the resulting
fast computations across many nodes? 

%By the time of writing this paper, we are aware of an going work for
%batching tasks in Spark~\cite{drizzle} to avoid the control overhead.
%The suggested solution requires loop-unrolling, and depends on
%programmer annotation for executing multiple batches. On the other
%hand, execution templates give granularity at the level of code blocks
%to support date dependent nested loops, and provide built in
%validation by design.

\medskip\noindent\textbf{Cloud schedulers} (also called cluster
managers) schedule tasks from many concurrent jobs across a collection
of worker nodes.  Because these schedulers have global knowledge of
all of the tasks in the system, they can efficiently multiplex jobs
across resources\cite{tetris}, improve job completion time~\cite{sjf},
fairly allocate resources across jobs~\cite{drf}, follow other
policies~\cite{apollo,tarcil,mesos}, or allow multiple algorithms to
operate on shared state~\cite{omega}.

% While initial cloud schedulers were centralized, increasing task rates
% from in-memory analytics workloads have led some systems to take a
% distributed or hybrid model.
Traditional centralized schedulers have transitioned to distributed or hybrid
models. 
In Sparrow~\cite{sparrow}, each job runs
its own independent scheduler that monitors the load on workers. These
schedulers independently make good cooperative scheduling decisions
based on mechanisms and principles derived from the power of two
choices~\cite{power-of-two}.  Tarcil uses a coarser grained approach,
in which multiple schedulers maintain copies of the full cluster
state, whose access is kept efficient through optimistic concurrency
control because conflicts are rare~\cite{tarcil}. Hawk's hybrid
approach centrally schedules long-running jobs for efficiency and
distributes short job scheduling for low latency~\cite{hawk}.
Finally, Mercury allows multiple schedulers to request resources
(``containers'') from a shared pool and then schedule tasks on their
resources~\cite{mercury}.

These distributed and hybrid schedulers address the problem of
when the combined task rate of multiple jobs is greater than what a
centralized scheduler can handle. Execution templates solve a similar,
but different problem, when the control plane bottlenecks a {\it single} 
job. Like Sparrow, a framework using execution templates
requests allocation from its cluster manager.

\medskip\noindent\textbf{High performance computing (HPC)} embraces
the idea that an application should be responsible for its own
scheduling as it has the greatest knowledge about its own performance
and behavior. HPC systems stretch from very low-level interfaces, such
as MPI~\cite{mpi}, which is effectively a high performance messaging
layer with some support for common operations such as reduction.
Partitioning and scheduling, however, is completely an application
decision, and MPI provides very little support for load balancing or
fault recovery. HPC frameworks such as Charm++~\cite{charmpp} and
Legion~\cite{legion} provide powerful abstractions to decouple control
flow, computation and communication, similar to cloud frameworks.
Their fundamental difference, however, is that these HPC systems only
provide mechanisms; applications are expected to provide their own
policies.

\section{Discussion and Conclusion}
\label{sec:discuss}

Analytics frameworks today provide either fine-grained scheduling or
high task throughput but not both. Execution templates enable a
framework to provide both simultaneously. By caching task graphs on
the controller and workers, execution templates are able to schedule
half a million tasks per second (Table~\ref{tab:instantiation-costs}). At
the same time, controllers can cheaply edit templates in response to
scheduling changes (Table~\ref{tab:edit-costs}). Finally, patches allow
execution templates to support dynamic control flow.

% Execution templates assume that a driver program repeats the same
% basic blocks many times.  For this reason, they are not valuable for
% short-lived jobs. Streaming computations, however, can benefit from
% standing templates that a driver periodically triggers. Execution
% templates also assume that a job's schedule, while dynamic, is mostly
% stable: Figure~\ref{fig:multi-tenant} showed that switching between
% cached schedules induces significant validation costs. If each
% iteration requires large edits or new templates, caching control
% decisions will not help.  This erratic behavior is rare, however,
% because such erratic schedules require large inter-worker data
% transfers.

Execution templates are a general control plane abstraction. However,
the requirements listed in Section~\ref{sec:design} are simpler to
incorporate in some systems than others.  Incorporating execution
templates into Spark requires two changes: workers need to queue tasks
and resolving dependencies locally and workers need to be able to
exchange data directly (not go through the controller for lookups).
Naiad's data flow graphs as well as TensorFlow's can be thought of as an extreme case of
execution templates, in which the flow graph describes a very large,
long-running basic block. Allowing a driver to store multiple graphs,
edit them, and dynamically trigger them
would bring most of the benefits.

\bibliography{nimbus}
\bibliographystyle{abbrv} 

\end{document}